\documentstyle[editedvolume]{crckapb}

\input epsf.sty
 
\def\etal{et al.~}

\def\Mbh{{M_{\bullet}}}
\def\Msph{{M_{\rm sph}}}
\def\Lsph{{L_{\rm sph}}}
\def\sigr{{\sigma_{\rm r}}}
\def\sigt{{\sigma_{\rm t}}}

\def\kms{{\rm\,km\,s^{-1}}}
\def\Mpc{{\rm\,Mpc}}

\def\AO{\bf [}
\def\AC{]\rm}
\def\GO{\sl [}
\def\GC{]\rm}

\begin{opening}
\title{Relics of Nuclear Activity:\protect\\ 
       do all galaxies have massive black holes?}
\subtitle{Invited Review for `Galaxy Interactions at Low and High 
          Redshift',\protect\\
          Proceedings IAU Symposium~186, Kyoto, August 1997,\protect\\ 
          D.~B.~Sanders, J.~Barnes, eds., Kluwer Academic Publishers.}
\author{Roeland P.~van der Marel}
\institute{STScI, 3700 San Martin Drive, Baltimore, MD 21218, USA}
\end{opening}

\runningtitle{Relics of Nuclear Activity}

\begin{document}

\vskip-0.5truecm
\begin{abstract}
The distribution of black hole (BH) masses $\Mbh$ in galaxies is
constrained by photometric and kinematic studies of individual
galaxies, and by the properties of the quasar population. I review our
understanding of these topics, present new results of adiabatic BH
growth models for HST photometry of elliptical galaxies with
brightness profiles of the `core' type, and discuss the implications
of ground-based stellar kinematical data. It is not yet possible to
uniquely determine the BH mass distribution, but the available
evidence is not inconsistent with a picture in which: (i) a majority
of galaxies has BHs; (ii) there is a correlation (with large scatter)
between $\Mbh$ and spheroid luminosity $\Lsph$ of the form $\Mbh
\approx 10^{-2} \Lsph$ (solar B-band units); and (iii) the BHs formed 
in a quasar phase through mass accretion with efficiency $\epsilon
\approx 0.05$.
\end{abstract}
\vskip-0.5truecm

\section{Introduction}

Considerable evidence suggests that the energetic processes in active
galaxies and quasars are due to the accretion of matter onto massive
BHs. Lynden-Bell (1969) already suggested that BHs may also be present
in quiescent galaxies, such as the Milky Way, M31 and M32. This
spurred efforts to find BHs in nearby galaxies through kinematical
studies, which have since increased steadily in sophistication, both
observationally and theoretically. There are now convincing BH
detections for at least a dozen galaxies, and new detections are
reported at an ever increasing rate. The techniques for detecting BHs
in individual galaxies have been reviewed by, e.g., Kormendy \&
Richstone (1995, hereafter KR95), Ford \etal (1998) and Richstone
(1998). Here I address the more general question: {\it do all galaxies
have BHs?}\looseness=-2

\section{Quasar counts and evolution}

Integration of quasar number counts yields the comoving energy density
in quasar light. Assuming that this energy is produced by accretion
onto massive black holes, one obtains the total mass per cubic Mpc
that is collected in black holes (So{\l}tan 1982). Division by the
observed luminosity density of galaxies (Loveday \etal 1992) yields an
estimate of the average black hole mass per unit luminosity: $\langle
\Mbh \rangle / \langle L \rangle = 2.0 \times 10^{-3} (0.1/\epsilon)$,
where $\epsilon$ is the accretion efficiency (Chokshi \& Turner
1992). [Throughout this paper, $H_0 = 80 \kms \Mpc^{-1}$,
mass-to-light ratios are in solar units, and luminosities are in the
B-band.]

To address the BH mass {\it distribution}, one must model not only the
total energy budget of the quasar population, but also its evolution.
Tremaine (1996; also Faber \etal 1997, hereafter F97) presented a
simple argument based on the typical quasar lifetime to show that a
model in which every spheroid (bulge or elliptical) has a BH is
consistent with the inferred $\langle \Mbh \rangle / \langle L
\rangle$. Haehnelt \& Rees (1993, hereafter HR93) presented a more
detailed model (in which BH formation is linked to hierarchical
structure formation) to fit the distribution of quasars as function of
magnitude and redshift. Their predicted BH mass distribution at the
current time (their Fig.~8) is consistent with a fraction $f \approx
0.3$ of all galaxies having a BH.\looseness=-2

The uncertainties in these estimates are considerable. The only
conclusion that can be drawn with some confidence is that a fraction
$f=0.1$--1 of all galaxies is likely to contain BHs with $\Mbh/L =
10^{-2}$--$10^{-3}$. The product of these quantities, $\langle \Mbh
\rangle / \langle L \rangle$, is better constrained than either
quantity independently, but is still rather uncertain (if only due to
the unknown~$\epsilon$).\looseness=-2

\begin{figure}
\noindent\begin{minipage}[b]{4.6truecm}
\epsfysize=4.65truecm
\epsfbox{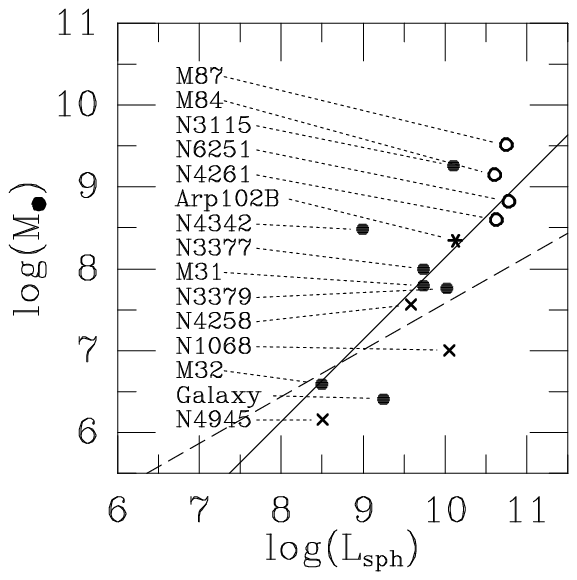}
\end{minipage}
\hfill
\noindent\begin{minipage}[b]{4.1truecm}
\epsfysize=4.65truecm
\epsfbox{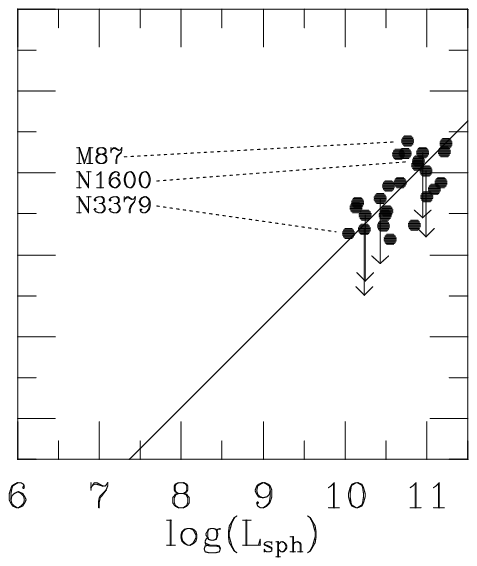}
\end{minipage}
\hfill
\noindent\begin{minipage}[b]{4.1truecm}
\epsfysize=4.65truecm
\epsfbox{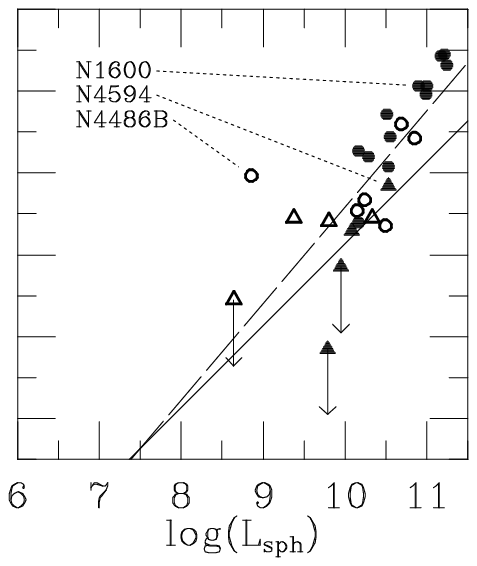}
\end{minipage}
\vskip-0.2truecm
\caption{Measurements of black hole mass $\Mbh$ versus spheroid 
luminosity $\Lsph$. Arrows indicate upper limits. {\bf (a;~left)}
`Secure' BH detections obtained with various techniques, as indicated
by different symbols and discussed in {\S}3. References are included
in the bibliography; where necessary, spheroid/total luminosity ratios
were estimated from the Hubble type and the relation in Simien \& de
Vaucouleurs (1986). {\bf (b;~middle)} $\Mbh$ values for all core
galaxies in F97 with $M_B \!<\! -20$, inferred from models of
adiabatic BH growth for HST photometry ({\S}4; van der Marel 1998, in
preparation). M87 and NGC 3379 are the two core galaxies for which
secure kinematic determinations are also available (cf.~panel~a). NGC
1600 is discussed in {\S}5. {\bf (c;~right)} $\Mbh$ determinations
from axisymmetric $f(E,L_z)$ models for ground-based stellar
kinematical observations of elliptical galaxies (Mg98). Circles:
`core' galaxies; triangles: `power-law' galaxies; open symbols:
galaxies in Virgo. Galaxies already included in the left panel are
omitted. For NGC 4486B and 4594 the presence of a BH has been
suggested previously on the basis of isotropic models. The $\Mbh$
values in panel~a are typically believed to have $|\Delta \log
\Mbh | \leq 0.3$. The accuracy of the $\Mbh$ measurements in panels~b
and~c is dominated by systematic uncertainties in the model
assumptions, as discussed in {\S}4, 5 and~6. In particular, mild
radial velocity anisotropy would lower the values in panel~c. The
solid line in each panel is the `reference model' discussed in the
{\S}3. It has $\Mbh = 1.4 \times 10^{-2}
\Lsph$, and fits the constraints from quasar number counts for an
assumed accretion efficiency $\epsilon=0.05$.  The long-dashed line in
the rightmost panel is an alternative model. It has $\Mbh = 6.0 \times
10^{-3} \Msph$ (with $\Msph \propto \Lsph^{1.18}$, cf.~Mg98), and fits
the constraints from quasar number counts for $\epsilon=0.022$. The
dashed line in the left panel indicates those $\Mbh$ for which
$r_{\bullet} = 0.1''$ at $D=10 \Mpc$.\looseness=-2}
\vskip-0.5truecm
\end{figure}

\section{BH detections}

HR93 predict a direct connection between $\Mbh$ and the galaxy
formation redshift $z_{\rm form}$ (galaxies that form later have
smaller BHs), but not between $M_{\bullet}$ and galaxy luminosity.
Spirals form later than ellipticals, and are therefore predicted to
have smaller BHs. Observations appear to confirm this; e.g., the
(active) galaxies M87 and NGC 1068 have similar luminosities, but
$\Mbh$ is $10^{2.5}$ times larger in M87 (cf.~Figure~1a
below). Observations do not rule out a correlation between $\Mbh$ and
the {\it spheroid} luminosity of the host (KR95). This may indicate
that $\Mbh$ and $\Lsph$ depend similarly on a common underlying
parameter (e.g., $z_{\rm form}$, as in the models of HR93), or
alternatively, that there is a physical link between BHs and
spheroids.

Figure~1a shows $\Mbh$ versus $\Lsph$ for all currently available BH
mass determinations inferred from: ($\times$)~radio observations of
water masers; ($\circ$)~ionized gas kinematics of nuclear disks;
($\ast$)~time variability of broad double-peaked Balmer lines; and
($\bullet$)~stellar kinematical studies that included aniso\-tropic
modeling (studies with only isotropic models are discussed in
{\S}6). There is indeed a correlation, but the scatter is large
($\sim\!2$ dex at fixed $\Lsph$) and selection bias may be important.
The dashed line shows the $\Mbh$ for which the BH sphere of influence,
$r_{\bullet} \simeq G \Mbh / \sigma^2$, extends $0.1''$ at a distance
$D = 10 \Mpc$ ($\sigma$ is determined by $\Lsph$ through the
Faber-Jackson relation). BHs below this line can be detected only in
galaxies closer than $10 \Mpc$, and in galaxies in which kinematical
tracers can be observed at resolutions $< 0.1''$ (e.g., water masers).

The solid line shows the predictions of one possible model that is
consistent with the $\langle \Mbh \rangle / \langle L \rangle$ from
quasar number counts. This `reference model' assumes that every
spheroid has a BH with $\Mbh \propto \Lsph$. Approximately 30\% of the
light from galaxies is due to spheroids (Schechter \& Dressler 1987).
So for an assumed accretion efficiency $\epsilon = 0.05$ this yields
$\Mbh = 1.4 \times 10^{-2} \Lsph$, which reproduces the trend in the
data.

\section{Surface-brightness profiles}

HST observations of early-type galaxies have revealed central surface
bright\-ness cusps that fall in two categories (F97), `power-laws'
(showing no clear break) and `cores' (showing a clear break). Cusps
can be explained as a consequence of the influence of a BH on
surrounding stars (Young 1980, hereafter Y80). Properties of cusps
around BHs depend on $\Mbh$, initial conditions (Quinlan \etal 1995,
hereafter Q95) and two-body relaxation (Bahcall \& Wolf 1976). Cusps
may also be due to processes unrelated to BHs (KR95), and they can
also be destroyed (Quinlan \& Hernquist 1997). Observed cusps
therefore do not uniquely constrain the BH masses in galaxies.

Nonetheless, simple models of adiabatic BH growth for observed
photometry of M87 (Young \etal 1978; Lauer \etal 1992; Crane \etal
1993) and several other galaxies imply BH masses that agree well with
kinematic determinations.  I have therefore started a study of
adiabatic BH growth models for a large sample of galaxies with
published HST photometry. These models may be particularly relevant
for core galaxies, for which the observed break in the brightness
profile may be associated with an originally homogeneous core. I have
used the software of Q95 to fit the photometric models of Y80 to all
core galaxies in the sample of F97 with $M_B \!<\!- 20$ (van der Marel
1998, in preparation). The models fit well in the central few arcsec
(RMS residual $\sim\!0.05$ mag/arcsec$^2$), and the photometrically
inferred $\Mbh$ appear meaningful: the kinematically determined $\Mbh$
for M87 and NGC 3379 (see Figure~1a) are reproduced to within $0.12$
and $0.50$ dex, respectively.\looseness=-2

Figure~1b shows the results for the whole sample. The $\Mbh$ are
remarkably consistent with the kinematical detections in Figure~1a,
and show a similar trend with $\Lsph$. So despite their simplicity, it
may well be that the Y80 models capture the essence of surface
brightness cusps in core galaxies. The prevalence of these cusps would
then imply that most or all core galaxies have BHs, with $\Mbh \propto
\Lsph$ as suggested by Figure~1b.

\section{Stellar kinematics and velocity dispersion anisotropy}

Stellar motions often provide the only kinematical tool to study BH
masses in quiescent galaxies, but the well-known degeneracy between
$\Mbh$ and velocity dispersion anisotropy (Binney \& Mamon 1982) is
still a major complication. This degeneracy can be resolved when high
resolution HST data are available (e.g., van der Marel \etal 1997;
Gebhardt \etal 1998), but such data are not yet available for many
galaxies. Lower resolution ground-based data are plentiful, but more
ambiguous to interpret. I use the case of NGC 1600, an E3 core galaxy
with no significant rotation, to illustrate this.\looseness=-2

Ground-based kinematical data with $\sim\!2''$ resolution
(Jedrzejewski \& Schechter 1989) show a mildly peaked velocity
dispersion profile, and HST photometry shows a shallow (F97),
marginally significant (Byun \etal 1996; Gebhardt \etal 1996), surface
brightness cusp. I construct spherical dynamical models (adequate for
the present purpose) following the approach of van der Marel (1994,
hereafter vdM94). I solve the Jeans equation for a given velocity
anisotropy profile profile, $\sigr/\sigt(r)$ (where $2\sigt^2
\!\equiv\! \sigma_{\theta}^2 + \sigma_{\phi}^2$), project and 
convolve the results, and compare with the data in a $\chi^2$ sense.
The normalization of the dispersion profile is determined by the
stellar mass-to-light ratio $\Upsilon$, and its shape is determined by
$\sigr/\sigt$ and $\Mbh / \Upsilon$.

\begin{figure}
\vskip-0.5truecm
\epsfxsize=12.5truecm
\epsfbox{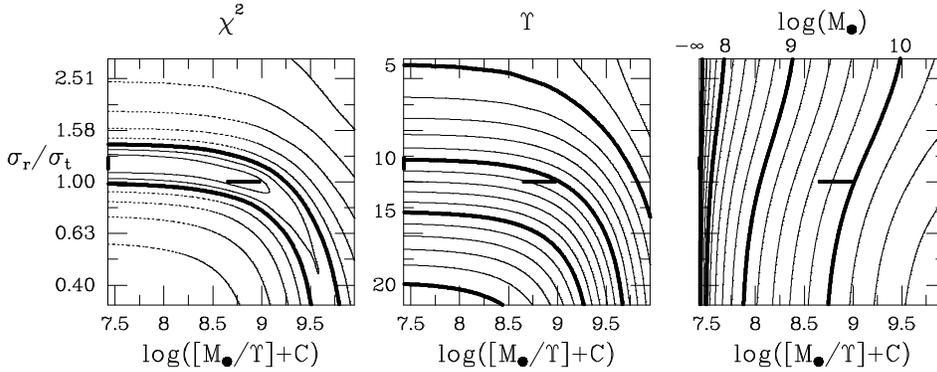}
\vskip-0.2truecm
\caption{Predictions of spherical dynamical models for NGC 1600 that
reproduce HST and ground-based photometry (F97; Peletier
\etal 1990). The abscissa in each panel is $\log([\Mbh/\Upsilon]+C)$,
where $\log C \equiv 7.425$; i.e., approximately logarithmic in
$(\Mbh/\Upsilon)$ for $(\Mbh/\Upsilon) \gg C$, and $\Mbh=0$ on the
left boundary. The ordinate is $\sigr/\sigt$, increasing
logarithmically from $1/3$ to $3$. For each
$(\Mbh/\Upsilon;\sigr/\sigt)$, the stellar mass-to-light ratio
$\Upsilon$ was chosen to minimize the $\chi^2$ of the fit to the
kinematical data of Jedrzejewski \& Schechter (1989). {\bf (a; left)}
Contours of $\chi^2$. The first two contours correspond to the
1$\sigma$ and 2$\sigma$ level, as calculated from the statistic
$\Delta \chi^2$. The heavy contours show the 3$\sigma$ region. {\bf
(b; middle)} Contours of $\Upsilon$, increasing linearly. {\bf (c;
right)} Contours of $\Mbh$, increasing logarithmically. Heavy curves
in panels~b and~c indicate contours for which the value of $\Upsilon$
or $\log \Mbh$ is indicated. Contours of $\Upsilon$ and $\chi^2$ are
approximately parallel, so $\Upsilon$ is well determined independent
of the anisotropy, $\log \Upsilon = 1.02$. By contrast, contours of
$\Mbh$ and $\chi^2$ are approximately perpendicular, so there is a
strong degeneracy between $\Mbh$ and anisotropy. In all panels, a
heavy horizontal bar indicates those isotropic models with a BH that
are acceptable at the 1$\sigma$ level, and a heavy vertical bar
indicates those anisotropic models without a BH that are acceptable at
the 1$\sigma$ level.}
\vskip-0.5truecm
\end{figure}

Figure~2 shows the fitted $\chi^2$, $\Mbh$ and $\Upsilon$ in the
$(\Mbh/\Upsilon;\sigr/\sigt)$ parameter space, for models with
constant $\sigr/\sigt$. The valley in the $\chi^2$ contours shows that
a one-parameter family of $(\Mbh/\Upsilon;\sigr/\sigt)$ combinations
fits the data. An isotropic model requires a very massive BH, $\log
\Mbh = 9.85$, but only modest anisotropy, $\sigr/\sigt = 1.18$, is
required to fit the data without a BH. Models that are radially
anisotropic at all radii may not correspond to a positive phase-space
distribution function (DF) (vdM94), or may not be stable (Stiavelli
\etal 1993), so I also constructed models in which the core is
isotropic, and in which there is a smooth transition at the break
radius of the surface brightness profile ($r_{\rm b} = 3.12''$) to a
value $(\sigr/\sigt)_{\rm main}$ characteristic of the main
body. Models with $\log \Mbh \!<\! 10.01$ are still all acceptable at
the 1$\sigma$ level. The best fit with no BH has $(\sigr/\sigt)_{\rm
main} = 1.45$, and the best fit with $\log
\Mbh = 9.15$, as suggested by adiabatic BH growth (Figure~1b), has
$(\sigr/\sigt)_{\rm main} = 1.30$. So in the absence of independent
constraints on the velocity dispersion anisotropy, the data do not
significantly constrain $\Mbh$.

Our understanding of the velocity anisotropy in galaxies is only
rudimentary. Ellipticals and bulges with power-law brightness profiles
have low to intermediate luminosity (F97), and are flattened by
rotation. The tensor virial theorem indicates that they may be
isotropic. Any anisotropic model with the same average $(\sigma_R^2 +
\sigma_{\phi}^2)/\sigma_z^2$ is also viable, but both M32 (van der Marel 
\etal 1998) and the Galactic Bulge (Evans \& de Zeeuw 1994) have indeed
been shown to be nearly isotropic. By contrast, core galaxies like NGC
1600 have intermediate to high luminosity, and little rotation (F97).
Detailed studies of three individual galaxies (Merritt \& Oh 1997; Rix
\etal 1997; Gerhard \etal 1998) are consistent with mild radial
anisotropy, $\sigr/\sigt = 1.2$--$1.4$, with a possible transition to
isotropy at small radii. Such a velocity distribution can be produced
by dissipationless collapse (van Albada 1982).  Mild radial anisotropy
is also consistent with studies of the ratio of major to minor axis
kinematics (van der Marel 1991) and line-of-sight velocity profile
shapes (Bender \etal 1994) in a larger sample of core galaxies, and is
also seen in the Galactic halo in the solar neighborhood ($\sigr/\sigt
= 1.5 \pm 0.2$; Beers \& Sommer-Larsen 1995). So it appears that
power-law galaxies may be approximately isotropic and that core galaxy
may be mildly radially anisotropic, but neither result is firmly
established.\looseness=-2

\section{Isotropic models for stellar kinematical data}

Magorrian \etal (1998, hereafter Mg98) studied 36 (mostly) elliptical
galaxies for which HST photometry and ground-based stellar kinematics
have been published. Each galaxy was modeled with the Jeans equations,
assuming an $f(E,L_z)$ DF (the axisymmetric generalization of a
spherical isotropic model). Figure~1c shows the BH masses that best
fit the observed kinematics. BHs are required in nearly all galaxies,
with $\Mbh \approx 6 \times 10^{-3} \Msph$ (long-dashed line),
consistent with quasar number counts if the accretion efficiency
$\epsilon = 0.022$. This is the first dynamical study that addresses a
large sample in a homogeneous way while including HST photometry.  It
establishes the important fact that the presence of a BH in every
spheroid is consistent with kinematical data, and that the required BH
masses are consistent with quasar counts for a reasonable value of
$\epsilon$.

Nonetheless, the Mg98 results are not unique. Of the 29 galaxies that
require a BH under the $f(E,L_z)$ hypothesis, 19 are core galaxies
with similar data as for NGC 1600. Mg98 find $\log \Mbh = 10.07$ for
NGC 1600, but the results in {\S}5 showed that all $\Mbh$ smaller than
this are equally acceptable. So the Mg98 models may have overestimated
the masses and/or prevalence of BHs. This would not violate the
constraints from quasar counts: if one assumes a higher
$\epsilon=0.1$, one may decrease all $\Mbh$ by a factor $4.5$, or
remove the BHs in 78\% of the galaxies.

Two core galaxies in the sample have $\Mbh$ determinations from
independent sources. Neither is well fit by an $f(E,L_z)$ model. For
M87, Mg98 infer the same $\Mbh$ as inferred from HST gas kinematics,
but only if the data outside $5''$ are ignored. For NGC 3379, Mg98
infer an $\Mbh$ that exceeds the more accurate determination of
Gebhardt \etal (1998) by a factor~7. Independent of whether one views
these comparisons as reasonable or poor agreement, it leaves open the
question whether $f(E,L_z)$ models return the correct result for
galaxies that may not have a (significant) BH.

One may wonder whether the correlation between $\Mbh$ and $\Lsph$
inferred by Mg98 can be explained if the $\Mbh$ values were partly
spurious. This is in fact the case. For galaxies that are radially
anisotropic, isotropic models will fit the observed dispersion
gradients by invoking BHs for which $r_{\bullet} \equiv G \Mbh /
\sigma^2$ is similar to the observational resolution. This predicts a
correlation of $\Mbh$ with distance of the form $r_{\bullet} \approx
2''$, which is not inconsistent with the Mg98 results. The more
distant galaxies in the sample are the most luminous. So this predicts
not only the correlation of $\Mbh$ with $\Lsph$, but also that this
correlation should be weaker for the galaxies in Virgo (which are all
at the same distance), as seen in Figure~1c.

Actual measurements of the velocity anisotropy are required to
establish whether or not the $\Mbh$ inferred by Mg98 are
correct. Either way, the $\Mbh$ in Figure~1c are 4.5 times higher than
those in Figure~1b, averaged over the 14 galaxies common to both
samples. So either the photometric measurements are too low (not
impossible, cf. the uncertainties discussed in~{\S}4), or the Mg98
results are too high (which would require mild radial anisotropy that
is not inconsistent with our understanding of core galaxies,
cf.~{\S}5).\looseness=-2

\section{Conclusions}

Our understanding of the BH mass distribution is still incomplete,
partly due to a lack of complete representative samples that cover
quiescent and active galaxies of all Hubble types, and partly due to
persistent uncertainties in the correct interpretation of photometric
and kinematic data. However, it is clear that we are finding BHs in
the correct mass range to explain quasar fueling and evolution, to
within the uncertainties.\looseness=-2

\vfill


\renewcommand{\baselinestretch}{1.0} 
\small 

I thank Gerry Quinlan for kindly allowing me to use his adiabatic BH
growth software. This work benefited from discussions with Eric
Emsellem, Tod Lauer, John Magorrian, Scott Tremaine and Tim de Zeeuw.
It was supported by STScI grant HF-1065.01-94A and an STScI
Fellowship. STScI is operated by AURA Inc., under NASA contract
NAS5-26555.

\newpage


\renewcommand{\baselinestretch}{1.0}


\end{document}